\documentclass[letterpaper,twocolumn,english,prl,nopreprint]{revtex4}
\usepackage[T1]{fontenc}
\usepackage[latin1]{inputenc}
\usepackage{amsmath}
\usepackage{babel}
\usepackage{graphics}
\usepackage{amssymb}

\makeatletter

\providecommand{\LyX}{L\kern-.1667em\lower.25em\hbox{Y}\kern-.125emX\@}

\newcommand{\C}[1]{{}}
\newcommand{\bS}{\begin{subequations}}
\newcommand{\eS}{\end{subequations}}
\newcommand{\sss}{\scriptscriptstyle}
\newcommand{\ie}{i.\,e.\ }
\newcommand{\eg}{e.\,g.\ }
\newcommand{\cf}{cf.\ }

\makeatother
\begin{document}

\newcommand{\para}{\mathsf{ll}}

\title{On the Symmetry of Universal Finite-Size Scaling Functions in Anisotropic
Systems}

\author{Alfred Hucht}

\affiliation{Theoretische Physik, Gerhard-Mercator-Universität, D-47048 Duisburg, Germany}

\email{fred@thp.Uni-Duisburg.de}

\date{\today}

\begin{abstract}
In this work a symmetry of universal finite-size scaling functions under
a certain anisotropic scale transformation is postulated. This transformation
connects the properties of a finite two-dimensional system at criticality
with generalized aspect ratio \( \rho >1 \) to a system with \( \rho <1 \).
The symmetry is formulated within a finite-size scaling theory, and expressions
for several universal amplitude ratios are derived. The predictions are
confirmed within the exactly solvable weakly anisotropic two-dimensional
Ising model and are checked within the two-dimensional dipolar in-plane
Ising model using Monte Carlo simulations. This model shows a strongly
anisotropic phase transition with different correlation length exponents
\( \nu _{\para }\neq \nu _{\perp } \) parallel and perpendicular to the
spin axis.
\end{abstract}


\maketitle
The theory of universal finite-size scaling (UFSS) functions is a key concept
in modern understanding of continuous phase transitions \cite{PrivmanFisher84,Privman90,HenkelSchollwoeck01}.
In particular, it is known that the UFSS functions of a rectangular two-dimensional
(2D) system of size \( L_{\para }\times L_{\perp } \) depend on the aspect
ratio \( L_{\para }/L_{\perp } \) \cite{Binder90}. For instance, in \emph{isotropic}
systems the scaling function at criticality \( \bar{U}_{\mathrm{c}} \)
of the Binder cumulant \( U=1-\frac{1}{3}\langle m^{4}\rangle /\langle m^{2}\rangle ^{2} \)
\cite{Binder81}, where \( \langle m^{n}\rangle  \) is the \( n \)-th
moment of the order parameter, is known to be a universal function \( \bar{U}_{\mathrm{c}}(L_{\para }/L_{\perp }) \)
for a given boundary condition. This quantity has been investigated by
several authors in the isotropic 2D Ising model with periodic boundary
conditions \cite{BinderWang89,KamieniarzBloete93}, while the influence
of other boundary conditions on \( \bar{U}_{\mathrm{c}}(L_{\para }/L_{\perp }) \)
has recently been studied in Refs.~\cite{Okabe99,KanedaOkabe01}.

In \emph{weakly} \emph{anisotropic} systems, where the couplings are anisotropic
(\( J_{\para }\neq J_{\perp } \) in the 2D Ising case), the correlation
length of the infinite system in direction \( \mu =\para ,\perp  \) becomes
anisotropic and scales like \( \xi _{\mu }^{(\infty )}(t)\sim \hat{\xi }_{\mu }t^{-\nu } \)
near criticality. (\( t=(T-T_{\mathrm{c}})/T_{\mathrm{c}} \) is the reduced
temperature and we assume \( t>0 \) without loss of generality.) This
leads to a correlation length amplitude ratio \( \hat{\xi }_{\para }/\hat{\xi }_{\perp } \)
different from unity. The UFSS functions then depend on this ratio, \ie
\( \bar{U}_{\mathrm{c}}=\bar{U}_{\mathrm{c}}(L_{\para }/L_{\perp },\hat{\xi }_{\para }/\hat{\xi }_{\perp }) \).
However, isotropy can be restored asymptotically by an anisotropic scale
transformation, where all lengths are rescaled with the corresponding correlation
length amplitudes \( \hat{\xi }_{\mu } \) \cite{LandauSwendsen84,Indekeu86,Yurishchev97}.
Thus the UFSS functions depend on \( L_{\para }/L_{\perp } \) and \( \hat{\xi }_{\para }/\hat{\xi }_{\perp } \)
only through the \emph{reduced} aspect ratio \( (L_{\para }/\hat{\xi }_{\para })/(L_{\perp }/\hat{\xi }_{\perp }) \).

In \emph{strongly} \emph{anisotropic} systems both the amplitudes \( \hat{\xi }_{\mu } \)
as well as the correlation length exponents \( \nu _{\mu } \) are different
and the correlation length in direction \( \mu  \) scales like \begin{equation}
\label{e:xi(t)}
\xi _{\mu }^{(\infty )}(t)\sim \hat{\xi }_{\mu }t^{-\nu _{\mu }}.
\end{equation}
Examples for strongly anisotropic phase transitions are Lifshitz points
\cite{Hornreich75} as present in the anisotropic next nearest neighbor
Ising (ANNNI) model \cite{Selke92,DiehlShpot00,PleimlingHenkel01}, or
the non-equilibrium phase transition in the driven lattice gas model \cite{KatzLebowitzSpohn83,SchmittmannZia95}.
Furthermore, in dynamical systems one can identify the \( \para  \)--direction
with time and the \( \perp  \)--direction(s) with space \cite{Henkel99},
which in most cases give strongly anisotropic behavior.

Using the same arguments as above we conclude that UFSS functions of strongly
anisotropic systems depend on the \emph{generalized} reduced aspect ratio
(\cf \cite{BinderWang89}) \begin{equation}
\label{e:rhorxi}
\rho =L_{\para }L_{\perp }^{-\theta }/r_{\xi },\quad \textrm{with}\quad r_{\xi }=\hat{\xi }_{\para }\hat{\xi }_{\perp }^{-\theta }
\end{equation}
 being the \emph{generalized} correlation length amplitude ratio, and with
the anisotropy exponent \( \theta =\nu _{\para }/\nu _{\perp } \) \cite{Henkel99}.
Up to now no attempts have been made to describe the dependency of UFSS
functions like \( \bar{U}_{\mathrm{c}}(\rho ) \) on the shape \( \rho  \)
of strongly anisotropic systems. In particular, it is not known if the
anisotropy exponent \( \theta  \) can be determined from \( \bar{U}_{\mathrm{c}}(\rho ) \).
This problem is addressed in this work.

\begin{figure}
[t]

{\centering \resizebox*{7cm}{!}{\includegraphics{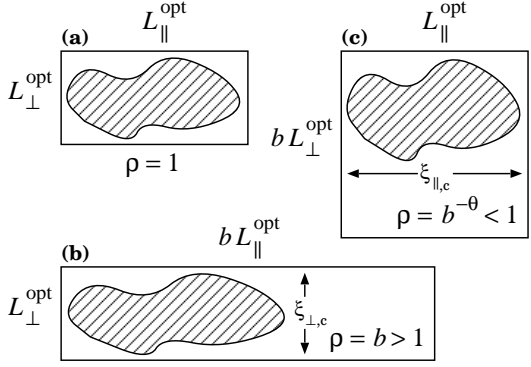}} \par}

\caption{Three systems with different aspect ratio \protect\( \rho \protect \)
(Eq.~(\ref{e:rhorxi})) at criticality. In (a) the critical correlation
volume \protect\( \xi _{\para ,\mathrm{c}}\xi _{\perp ,\mathrm{c}}\protect \)
(shaded area) spans the whole system, while in (b) and (c) correlations
are limited by symmetric finite-size effects. \label{f:systems}}
\end{figure}

Consider a 2D strongly anisotropic finite system with periodic boundary
conditions. When the critical point of the infinite system is approached
from temperatures \( t>0 \), the correlation lengths \( \xi _{\mu } \)
in the different directions \( \mu  \) are limited by the direction in
which \( \xi _{\mu }^{(\infty )} \) from Eq.~(\ref{e:xi(t)}) reaches
the system boundary first \cite{Binder90}. For a given volume \( N=L_{\para }L_{\perp } \)
we define an {}``optimal'' shape \( L_{\para }^{\mathrm{opt}}\times L_{\perp }^{\mathrm{opt}} \)
at which both correlation lengths \( \xi _{\mu }^{(\infty )} \) reaches
the system boundary simultaneously, \ie \begin{equation}
\label{e:Lopt}
L^{\mathrm{opt}}_{\mu }:=\xi ^{(\infty )}_{\mu }(t)
\end{equation}
 for some temperature \( t>0 \) (Fig.~\ref{f:systems}a). We immediately
find using Eqs.~(\ref{e:xi(t)}, \ref{e:rhorxi}) that the optimal shape
obeys \( \rho _{\mathrm{opt}}\equiv 1 \) for all \( N \), giving \( L_{\para }^{\mathrm{opt}}=r_{\xi }(L_{\perp }^{\mathrm{opt}})^{\theta } \).
A system of optimal shape should show the strongest critical fluctuations
for a given volume \( N \) as the critical correlation volume \( \xi _{\para ,\mathrm{c}}\xi _{\perp ,\mathrm{c}} \)
spans the whole system.

At the optimal aspect ratio \( \rho =1 \) the correlations are limited
by both directions \( \para  \) and \( \perp  \) (Fig.~\ref{f:systems}a).
If the system is enlarged by a factor \( b>1 \) in the \( \para  \)--direction
(Fig.~\ref{f:systems}b), the correlation volume may relax into this direction
but does not fill the whole system due to the limitation in \( \perp  \)--direction.
A similar situation with exchanged roles occurs if the system is enlarged
by a factor \( b>1 \) in the \( \perp  \)--direction (Fig.~\ref{f:systems}c).
We now \emph{assume} that systems (b) and (c) are similar in the scaling
region \( L_{\mu }^{\mathrm{opt}}\rightarrow \infty  \), \ie that their
correlation volumes are asymptotically equal.

Hence we can formulate a \emph{symmetry} \emph{hypothesis}: Consider a
system with periodic boundary conditions and optimal aspect ratio \( \rho =1 \)
at the critical point. If this system is enlarged by a factor \( b>1 \)
in \( \para  \)--direction, it behaves asymptotically the same as if enlarged
by \emph{the same factor} \( b \) in \( \perp  \)--direction. 

To formulate this hypothesis within a finite-size scaling theory, we consider
a 2D strongly anisotropic system of size \( L_{\para }\times L_{\perp } \)
which fulfills the generalized hyperscaling relation \( 2-\alpha =\nu _{\para }+\nu _{\perp } \)
\cite{BinderWang89}. For our purpose it is sufficient to focus on the
critical point. The universal finite-size scaling \emph{ansatz} \cite{PrivmanFisher84,BinderWang89,Privman90,Binder90,HenkelSchollwoeck01}
for the singular part of the free energy density \( f_{\mathrm{c}}=F_{\mathrm{s},\mathrm{c}}/(Nk_{\mathrm{B}}T_{\mathrm{c}}) \)
reads \cite{Hucht-inprep} \begin{equation}
\label{e:fY}
f_{\mathrm{c}}(L_{\para },L_{\perp })\sim \frac{b_{\para }b_{\perp }}{N}Y_{\mathrm{c}}(b_{\para },b_{\perp })
\end{equation}
 with the scaling variables \( b_{\mu }=\lambda ^{\nu _{\mu }}L_{\mu }/\hat{\xi }_{\mu } \),
where \( \lambda  \) is a free scaling parameter. The scaling function
\( Y_{\mathrm{c}} \) is universal for a given boundary condition, all
non-universal properties are contained in the metric factors \( \hat{\xi }_{\mu } \).
These metric factors occur due to the usual requirement that the relevant
lengths are \( L_{\mu }/\xi _{\mu }^{(\infty )}(t) \) near criticality,
and cannot be absorbed into \( \lambda  \) in contrast to isotropic systems. For
the three systems in Fig.~\ref{f:systems} we set \( \lambda =(L^{\mathrm{opt}}_{\mu }/\hat{\xi }_{\mu })^{-1/\nu _{\mu }} \)
to get \bS \label{e:fYabc}\begin{eqnarray}
f_{\mathrm{c}}(L^{\mathrm{opt}}_{\para },L^{\mathrm{opt}}_{\perp }) & \sim  & \frac{1}{N}Y_{\mathrm{c}}(1,1)\label{e:fYa} \\
f_{\mathrm{c}}(bL^{\mathrm{opt}}_{\para },L^{\mathrm{opt}}_{\perp }) & \sim  & \frac{b}{N}Y_{\mathrm{c}}(b,1)\label{e:fYb} \\
f_{\mathrm{c}}(L^{\mathrm{opt}}_{\para },bL^{\mathrm{opt}}_{\perp }) & \sim  & \frac{b}{N}Y_{\mathrm{c}}(1,b).\label{e:fYc} 
\end{eqnarray}
 \eS The proposed symmetry hypothesis states that for \( b>1 \) Eqs.~(\ref{e:fYb})
and (\ref{e:fYc}) are asymptotically equal in the scaling region where
\( L^{\mathrm{opt}}_{\mu } \) is large, \begin{equation}
\label{e:fsymm}
f_{\mathrm{c}}(bL^{\mathrm{opt}}_{\para },L^{\mathrm{opt}}_{\perp })\stackrel{\sss b>1}{\sim }f_{\mathrm{c}}(L^{\mathrm{opt}}_{\para },bL^{\mathrm{opt}}_{\perp }).
\end{equation}
 Hence the scaling function \( Y_{\mathrm{c}} \) has the simple symmetry
\begin{equation}
\label{e:Ysymm}
Y_{\mathrm{c}}(b,1)\stackrel{\sss b>1}{=}Y_{\mathrm{c}}(1,b).
\end{equation}
 To rewrite \( Y_{\mathrm{c}} \) as function of the generalized aspect
ratio \( \rho  \) (Eq.~(\ref{e:rhorxi})) instead of the quantities \( b_{\mu } \),
we set \( b_{\perp }=1 \) in system (c) and get, as then \( \lambda =(bL^{\mathrm{opt}}_{\perp }/\hat{\xi }_{\perp })^{-1/\nu _{\perp }} \),
\begin{equation}
\label{e:fYc2}
f_{\mathrm{c}}(L^{\mathrm{opt}}_{\para },bL^{\mathrm{opt}}_{\perp })\sim \frac{b^{-\theta }}{N}Y_{\mathrm{c}}(b^{-\theta },1).
\end{equation}
Eqs.~(\ref{e:fYc}) and (\ref{e:fYc2}) are identical and we conclude
that \( bY_{\mathrm{c}}(1,b)=b^{-\theta }Y_{\mathrm{c}}(b^{-\theta },1) \).
At this point it is convenient to define the scaling function \( \bar{Y}_{\mathrm{c}}(b)=bY_{\mathrm{c}}(b,1) \)
which fulfills \begin{equation}
\label{e:fYbar}
f_{\mathrm{c}}(L_{\para },L_{\perp })\sim \frac{1}{N}\bar{Y}_{\mathrm{c}}(\rho ).
\end{equation}
 For this scaling function the symmetry reads \begin{equation}
\label{e:fYbarsymm}
\bar{Y}_{\mathrm{c}}(\rho )\stackrel{\sss \rho >1}{=}\bar{Y}_{\mathrm{c}}(\rho ^{-\theta }).
\end{equation}
 We see from Eq.~(\ref{e:fYbar}) that the critical free energy density
\( f_{\mathrm{c}} \) is a universal function of the reduced aspect ratio
\( \rho =L_{\para }L_{\perp }^{-\theta }/r_{\xi } \) without any non-universal
prefactor, and that at criticality \emph{all} system specific properties
are contained in the non-universal ratio \( r_{\xi } \) from Eq.~(\ref{e:rhorxi}).

Ansatz Eq.~(\ref{e:fY}) can also be made for the inverse spin-spin correlation
length at criticality \cite{Hucht-inprep} \begin{equation}
\label{e:kX}
\xi ^{-1}_{\mu ,\mathrm{c}}(L_{\para },L_{\perp })\sim \frac{b_{\mu }}{L_{\mu }}X_{\mu ,\mathrm{c}}(b_{\para },b_{\perp }).
\end{equation}
 The proposed symmetry gives \( X_{\mu ,\mathrm{c}}(b,1)\stackrel{\sss b>1}{=}X_{\bar{\mu },\mathrm{c}}(1,b) \),
where \( \bar{\mu } \) denotes the direction perpendicular to \( \mu  \).
Hence the correlation volumes \( \xi _{\para ,\mathrm{c}}\xi _{\perp ,\mathrm{c}} \)
of system (b) and (c) in Fig.~\ref{f:systems} are indeed equal as assumed
above and become \( \xi _{\para ,\mathrm{c}}\xi _{\perp ,\mathrm{c}}\sim \frac{N}{b}X^{-1}_{\para ,\mathrm{c}}(b,1)X^{-1}_{\perp ,\mathrm{c}}(b,1) \). 

The correlation length amplitudes \( A_{\xi }^{\mu } \) in cylindrical
geometry (\( b_{\mu }\rightarrow \infty  \), \( b_{\bar{\mu }}=1 \)),
which can be calculated exactly for many isotropic two-dimensional models
within the theory of conformal invariance \cite{Cardy84} generalize to
the strongly anisotropic form \cite{HenkelSchollwoeck01} \begin{equation}
\label{e:Akdef}
A_{\xi }^{\mu }=\lim _{L_{\bar{\mu }}\rightarrow \infty }L_{\bar{\mu }}^{-\nu _{\mu }/\nu _{\bar{\mu }}}\lim _{L_{\mu }\rightarrow \infty }\xi _{\mu ,\mathrm{c}}(L_{\para },L_{\perp }).
\end{equation}
 Inserting Eq.~(\ref{e:kX}) they become \begin{equation}
\label{e:Ak}
A_{\xi }^{\para }=r_{\xi }X^{-1}_{\para ,\mathrm{c}}(\infty ,1),\quad A_{\xi }^{\perp }=r_{\xi }^{-1/\theta }X^{-1}_{\perp ,\mathrm{c}}(1,\infty )
\end{equation}
 which shows that in general \( A_{\xi }^{\mu } \) ist not universal.
The symmetry hypothesis states that both limits of the scaling function
\( X_{\mu ,\mathrm{c}} \) are equal. Denoting this universal limit \( A_{\xi }:=X^{-1}_{\para ,\mathrm{c}}(\infty ,1)=X^{-1}_{\perp ,\mathrm{c}}(1,\infty ) \)
we obtain \( A_{\xi }^{\para }=r_{\xi }A_{\xi } \) and \( A_{\xi }^{\perp }=r_{\xi }^{-1/\theta }A_{\xi } \)
as well as the amplitude relations \begin{equation}
\label{e:Akidentities}
A^{1+\theta }_{\xi }=A_{\xi }^{\para }(A_{\xi }^{\perp })^{\theta },\qquad \frac{A_{\xi }^{\para }}{A_{\xi }^{\perp }}=r_{\xi }^{1+1/\theta }.
\end{equation}

These predictions can be checked within the exactly solved weakly anisotropic
2D Ising model with different couplings \( J_{\para } \) and \( J_{\perp } \),
where the paramagnetic correlation length reads \( \xi _{\mu }^{(\infty )}(t)=(\log \coth (\beta J_{\mu })-2\beta J_{\bar{\mu }})^{-1} \)
with \( \beta =1/k_{\mathrm{B}}T \) \cite{Onsager44}. The amplitude ratio
\( r_{\xi } \) at the critical point \( \sinh (2\beta _{\mathrm{c}}J_{\para })\sinh (2\beta _{\mathrm{c}}J_{\perp })=1 \)
\cite{Onsager44} becomes \( r_{\xi }=\sinh (2\beta _{\mathrm{c}}J_{\para }) \)
\cite{ZiaAvron82}. On the other hand, the inverse correlation length amplitudes
in cylinder geometry Eq.~(\ref{e:Akdef}) has been calculated \cite{NightingaleBloete83}
to give \( A_{\xi }^{\mu }=\frac{4}{\pi }\sinh (2\beta _{\mathrm{c}}J_{\mu }), \)
which immediately yields Eqs.~(\ref{e:Ak}) if we insert the well known
universal value \( A_{\xi }=4/\pi  \) \cite{Luck82,Cardy84}. The left
relation of Eqs.~(\ref{e:Akidentities}) has already been derived for
several weakly anisotropic models, where it simplifies to \( A_{\xi }^{2}=A_{\xi }^{\para }A_{\xi }^{\perp } \)
\cite[Eq. (7)]{NightingaleBloete83}.

To check the symmetry numerically in strongly an\-isotropic systems, we
now focus on the Binder cumulant \( U \). The scaling \emph{ansatz} at
criticality Eq.~(\ref{e:fY}) becomes \begin{equation}
\label{e:UU}
U_{\mathrm{c}}(L_{\para },L_{\perp })\sim \frac{1}{b_{\para }b_{\perp }}\tilde{U}_{\mathrm{c}}(b_{\para },b_{\perp })=\bar{U}_{\mathrm{c}}(\rho )
\end{equation}
 with the scaling function \( \bar{U}_{\mathrm{c}}(b)=\tilde{U}_{\mathrm{c}}(b,1)/b \),
and the calculation is completely analogous to the free energy case. The
symmetry hypothesis for the cumulant scaling functions \( \tilde{U}_{\mathrm{c}} \)
and \( \bar{U}_{\mathrm{c}} \) thus reads (\cf Eqs.~(\ref{e:Ysymm},\ref{e:fYbarsymm}))
\begin{equation}
\label{e:Usymm}
\tilde{U}_{\mathrm{c}}(b,1)\stackrel{\sss b>1}{=}\tilde{U}_{\mathrm{c}}(1,b),\qquad \bar{U}_{\mathrm{c}}(\rho )\stackrel{\sss \rho >1}{=}\bar{U}_{\mathrm{c}}(\rho ^{-\theta }).
\end{equation}
The generalization of the cumulant amplitude \( A^{\mu }_{U} \) \cite{Binder81, BurkhardtDerrida85}
to strongly anisotropic systems is similar to Eq.~(\ref{e:Akdef}) and
gives \begin{equation}
\label{e:AUdef}
A_{U}^{\mu }=\lim _{L_{\bar{\mu }}\rightarrow \infty }L_{\bar{\mu }}^{-\nu _{\mu }/\nu _{\bar{\mu }}}\lim _{L_{\mu }\rightarrow \infty }L_{\mu }U_{\mathrm{c}}(L_{\para },L_{\perp }).
\end{equation}
 Inserting the scaling \emph{ansatz} Eq.~(\ref{e:UU}) we now find \begin{equation}
\label{e:AU}
A_{U}^{\para }=r_{\xi }\tilde{U}_{\mathrm{c}}(\infty ,1),\qquad A_{U}^{\perp }=r^{-1/\theta }_{\xi }\tilde{U}_{\mathrm{c}}(1,\infty ),
\end{equation}
 which again are in general not universal. Using the symmetry hypothesis
we can define \( A_{U}:=\tilde{U}_{\mathrm{c}}(\infty ,1)=\tilde{U}_{\mathrm{c}}(1,\infty ) \)
and get \( A_{U}^{\para }=r_{\xi }A_{U} \), \( A_{U}^{\perp }=r_{\xi }^{-1/\theta }A_{U} \)
as well as the identities (cf. Eqs.~(\ref{e:Akidentities}))\begin{equation}
\label{e:AUidentities}
A^{1+\theta }_{U}=A_{U}^{\para }(A_{U}^{\perp })^{\theta },\qquad \frac{A_{U}^{\para }}{A_{U}^{\perp }}=r_{\xi }^{1+1/\theta }.
\end{equation}

\begin{figure}
{\centering \resizebox*{8cm}{!}{\includegraphics{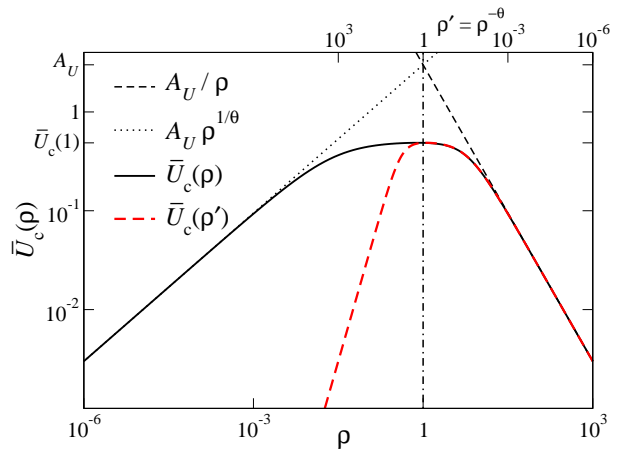}} \par}

\caption{Sketch of critical cumulant scaling functions \protect\( \bar{U}_{\mathrm{c}}(\rho )\protect \)
and \protect\( \bar{U}_{\mathrm{c}}(\rho ')\protect \) with \protect\( \rho '=\rho ^{-\theta }\protect \)
for assumed anisotropy exponent \protect\( \theta =2\protect \). We have
\protect\( \bar{U}_{\mathrm{c}}(\rho {\gg }1)\sim A_{U}/\rho \protect \)
and \protect\( \bar{U}_{\mathrm{c}}(\rho {\ll }1)\sim A_{U}\rho ^{1/\theta }\protect \).
For \protect\( \rho >1\protect \) \protect\( \bar{U}_{\mathrm{c}}(\rho )\protect \)
fulfills \protect\( \bar{U}_{\mathrm{c}}(\rho )=\bar{U}_{\mathrm{c}}(\rho ')\protect \).
\label{f:Uskizze}}
\end{figure}
The cumulant scaling function \( \bar{U}_{\mathrm{c}}(\rho ) \) must be
extremal at \( \rho =1 \) due to symmetry. Furthermore, as a deviation
from the optimal aspect ratio \( \rho =1 \) reduces the cumulant, it has
a maximum at this point \cite{BinderWang89}. A sketch of \( \bar{U}_{\mathrm{c}}(\rho ) \)
for an assumed anisotropy exponent \( \theta =2 \) is depicted in Fig.~\ref{f:Uskizze}.
For \( \rho >1 \) both \( \bar{U}_{\mathrm{c}}(\rho ) \) and \( \bar{U}_{\mathrm{c}}(\rho '=\rho ^{-\theta }) \)
collapse onto a single curve, reflecting the proposed symmetry. It is obvious
from Fig.~\ref{f:Uskizze} that \( \bar{U}_{\mathrm{c}}(\rho ) \) (and
thus also \( \bar{Y}_{\mathrm{c}}(\rho ) \) from Eq.~(\ref{e:fYbarsymm}))
can not be analytic at \( \rho =1 \) in strongly anisotropic systems,
as the two branches \( \bar{U}_{\mathrm{c}}(\rho ) \) and \( \bar{U}_{\mathrm{c}}(\rho ') \)
identical for \( \rho >1 \) fork at \( \rho =1 \) \cite{Hucht-inprep}.
On the other hand, \( \bar{Y}_{\mathrm{c}}(\rho ) \) and \( \bar{U}_{\mathrm{c}}(\rho ) \)
can be analytic at \( \rho =1 \) if the anisotropy exponent \( \theta =1 \),
as in the case of the isotropic 2D Ising model \cite[Eq. 3.37]{FerdinandFisher69}.

To check the symmetry hypothesis in a strongly anisotropic system, I performed
Monte Carlo simulations of the two-dimensional dipolar in-plane Ising model
\cite{Hucht-inprep} \begin{equation}
\label{e:H}
\mathcal{H}=-\frac{J}{2}\sum _{\langle ij\rangle }\sigma _{i}\sigma _{j}+\frac{\omega }{2}\sum _{i\neq j}\frac{(r^{\perp }_{ij})^{2}-2(r^{\para }_{ij})^{2}}{|\vec{r}_{ij}|^{5}}\sigma _{i}\sigma _{j}
\end{equation}
with spin variables \( \sigma =\pm 1 \), ferromagnetic nearest neighbor
exchange interaction \( J>0 \), and dipole interaction \( \omega >0 \).
The distance \( \vec{r}_{ij}=(r_{ij}^{\para },r_{ij}^{\perp }) \) between
spin \( \sigma _{i} \) and \( \sigma _{j} \) is decomposed into contributions
parallel and perpendicular to the spin axis. In the simulations the Wolff
cluster algorithm \cite{Wolff89} for long range systems proposed by Luijten
and Blöte \cite{LuijtenBloete95} was used, modified to anisotropic interactions.
In contrast to earlier work \cite{DeBell89,Bulenda00} using renormalization
group technics it is found that this model shows a strongly anisotropic
phase transition. The details of the simulations will be published elsewhere
\cite{Hucht-inprep}.

\begin{figure}
{\centering \resizebox*{8cm}{!}{\includegraphics{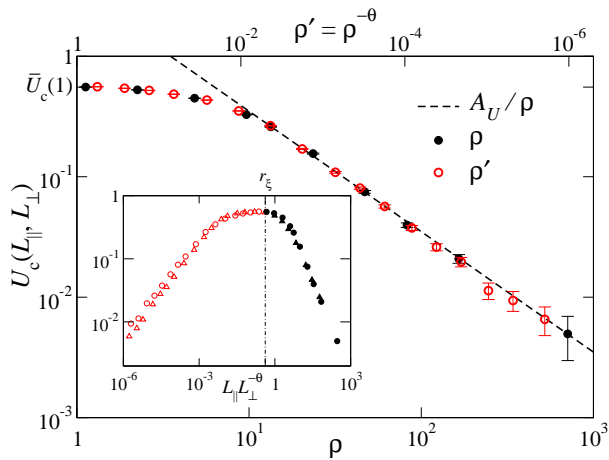}} \par}

\caption{Cumulant \protect\( U_{\mathrm{c}}(L_{\para },L_{\perp })\protect \)
of the dipolar in-plane Ising model (Eq.~(\ref{e:H})) for dipole strength
\protect\( \omega /J=0.1\protect \) and system size \protect\( N=43200\protect \)
at the critical point \protect\( k_{\mathrm{B}}T_{\mathrm{c}}/J=2.764(1)\protect \).
The data points collapse for \protect\( \rho >1\protect \) if we set \protect\( \theta =2.1(3)\protect \)
and \protect\( r_{\xi }=0.415(40)\protect \), giving the universal amplitudes
\protect\( \bar{U}_{\mathrm{c}}(1)=0.555(5)\protect \) and \protect\( A_{U}=3.5(2)\protect \).
The inset shows \protect\( U_{\mathrm{c}}\protect \) as function of the
un-reduced generalized aspect ratio \protect\( L_{\para }L^{-\theta }_{\perp }\protect \)
for system size \protect\( N=43200\protect \) (circles) and \protect\( N=4320\protect \)
(triangles). \label{f:Ucollapse}}
\end{figure}
After \( T_{\mathrm{c}} \) was determined, systems with constant volume
\( N=L_{\para }L_{\perp } \) were simulated, which was chosen to have
a large number of divisors in order to get many different aspect ratios
(\eg \( N=2^{6}3^{3}5^{2}=43200 \) has \( 84 \) divisors). The resulting
critical cumulant \( U_{\mathrm{c}}(L_{\para }L^{-\theta }_{\perp }) \)
for two different volumes \( N=4320,43200 \) is depicted in the inset
of Fig.~\ref{f:Ucollapse}. As expected, both curves have the same maximum
value \( \bar{U}_{\mathrm{c}}(1)=0.555(5) \) at criticality. With variation
of \( \theta  \) the curves are shifted horizontally and collapse for
\( \theta =2.1(3) \), with maximum at \( r_{\xi }=0.415(40) \). To check
the proposed symmetry we fold the left branch with \( \rho <1 \) (open
symbols) to the right and rescale the \( \rho  \)-axis with \( \theta  \).
The resulting data collapse for \( \rho >1 \) is shown in Fig.~\ref{f:Ucollapse}.
This collapse and the additional condition that both curves must go to
zero as \( A_{U}/\rho  \) allows a precise determination of \( \theta  \)
and \( r_{\xi } \) as well as of the universal amplitude \( A_{U}=3.5(2) \).

In conclusion, I postulate a symmetry of universal finite-size scaling
functions under a certain anisotropic scale transformation and generalize
the Privman-Fisher equations \cite{PrivmanFisher84} to strongly anisotropic
phase transitions on rectangular lattices at criticality. It turns out
that for a given boundary condition the only relevant variable is the generalized
reduced aspect ratio \( \rho =L_{\para }L_{\perp }^{\theta }/r_{\xi } \)
and that \eg the free energy scaling function Eq.~(\ref{e:fYbar}) obeys
the symmetry \( \bar{Y}_{\mathrm{c}}(\rho )\stackrel{\sss \rho >1}{=}\bar{Y}_{\mathrm{c}}(\rho ^{-\theta }) \).
At criticality, the free energy density \( f_{\mathrm{c}} \), the inverse
correlation lengths \( \xi _{\mu ,\mathrm{c}} \), and the Binder cumulant
\( U_{\mathrm{c}} \) are universal functions of \( \rho  \), without
a non-universal prefactor. All system specific properties are contained
in the non-universal correlation length amplitude ratio \( r_{\xi } \)
(Eq.~(\ref{e:rhorxi})).

The generalization to higher dimensions is straightforward \cite{Hucht-inprep},
an interesting application would be the precise determination of the exponent
\( \theta  \) at the Lifshitz point of the three-dimensional ANNNI model
\cite{DiehlShpot00,PleimlingHenkel01}. An open question is the validity
of the proposed symmetry in non-equilibrium systems with appropriate boundary
conditions, which recently have been shown to exhibit Privman-Fisher universality
\cite{HenkelSchollwoeck01}.

\begin{acknowledgments}
I thank Sven Lübeck and Erik Luijten for valuable discussions and Malte
Henkel for a critical reading of the manuscript. This work was supported
by the Deutsche Forschungsgemeinschaft through SFB~491.
\end{acknowledgments}
\bibliographystyle{apsrev}
\bibliography{Physik}

\end{document}